\documentclass[footinbib,aps,prl,reprint,superscriptaddress,nopacs]{revtex4-2}

\usepackage[absolute]{textpos}
\usepackage{physics, graphicx,amsmath,mathtools,comment}

\DeclareMathOperator{\sF}{\mathcal{F}}
\usepackage{lineno}

\begin{document}
\title{Building a controlled-NOT gate between polarization and frequency}

\author{Hsuan-Hao Lu}
\email{luh2@ornl.gov}
\affiliation{Quantum Information Science Section, Computational Sciences and Engineering Division, Oak Ridge National Laboratory, Oak Ridge, Tennessee 37831, USA}

\author{Joseph M. Lukens}
\affiliation{Quantum Information Science Section, Computational Sciences and Engineering Division, Oak Ridge National Laboratory, Oak Ridge, Tennessee 37831, USA}
\affiliation{Research Technology Office and Quantum Collaborative, Arizona State University, Tempe, Arizona 85287, USA}

\author{Muneer~Alshowkan}
\affiliation{Quantum Information Science Section, Computational Sciences and Engineering Division, Oak Ridge National Laboratory, Oak Ridge, Tennessee 37831, USA}

\author{Brian T. Kirby}
\affiliation{DEVCOM Army Research Laboratory, Adelphi, Maryland 20783, USA}
\affiliation{Tulane University, New Orleans, LA 70118, USA}

\author{Nicholas~A.~Peters}
\affiliation{Quantum Information Science Section, Computational Sciences and Engineering Division, Oak Ridge National Laboratory, Oak Ridge, Tennessee 37831, USA}

\date{\today}

\maketitle

\textbf{By harnessing multiple degrees of freedom (DoFs) within a single photon, controlled quantum unitaries, such as the two-qubit controlled-NOT (\textsc{cnot}) gate, play a pivotal role in advancing quantum communication protocols like dense coding and entanglement distillation. In this work, we devise and realize a \textsc{cnot} operation between polarization and frequency DoFs by exploiting directionally dependent electro-optic phase modulation within a fiber Sagnac loop. Alongside computational basis measurements, we validate the effectiveness of this operation through the synthesis of all four Bell states in a single photon, all with fidelities greater than 98\%. This demonstration opens new avenues for manipulating hyperentanglement across these two crucial DoFs, marking a foundational step toward leveraging polarization-frequency resources in fiber networks for future quantum applications.
}
\smallskip
\\
\textit{Introduction.---}Photons possess a rich array of degrees of freedom (DoFs) ripe for exploitation in various encoding schemes, facilitating information transmission across both free-space and fiber-optic networks. In response to escalating global bandwidth demands in \emph{classical} communications, 
a variety of multiplexing techniques---ranging from traditional wavelength-division multiplexing~\cite{Keiser1999} to more recent spatial multiplexing~\cite{Richardson2013}---are increasingly deployed to enhance transmission capacity~\cite{Agrell2016}. These multiplexing strategies also hold promise for future \emph{quantum} networks, encompassing multi-user entanglement distribution~\cite{Wengerowsky2018, Lingaraju2019, Alshowkan2021, Appas2021} and multiplexed quantum memories essential for future quantum repeater designs~\cite{Collins2007,Simon2007, Sangouard2011}. Nevertheless, in the exploration of multiple photonic DoFs, \emph{hyperentanglement} emerges as a distinct area of research, 
describing the simultaneous entanglement in multiple independent DoFs between two~\cite{Kwiat1997, Barreiro2005} or even more parties~\cite{Gao2010, Wang2018}.

The introduction of additional DoFs in photonic hyperentanglement naturally expands the dimension of the Hilbert space, thus unlocking new potential for quantum communication and networking, such as superdense coding~\cite{Barreiro2008} and teleportation~\cite{Bernstein2006, Graham2015, Chapman2020}, and single-copy entanglement distillation~\cite{Simon2002, Sheng2010, Hu2021, Ecker2021}. A notable departure lies in the design of two-qubit controlled logic: while traditional approaches operate on two separate particles (i.e., photons) using a single DoF, such operations can now occur within a single photon between two DoFs---one as the control qubit and the other as the target qubit---and enable deterministic functionality even with linear optics~\cite{Cerf1998, Fiorentino2004,Kagalwala2017,Reimer2019, Imany2019}. Such deterministic gates cannot supplant probabilistic multiphoton operations for universal quantum computing~\cite{Knill2001, Kok2007}, for the scaling is ultimately capped by the number of available DoFs. But they can prove extremely valuable in boosting the efficiency of few-qubit operations in quantum communications.

Among the DoFs available for exploration, polarization-frequency hyperentanglement stands out due to its favorable characteristics for state manipulation (polarization) and generation (frequency-bin, or in general, time-energy). This is evidenced by the proliferation of polarization-frequency hyperentangled sources in recent years, spanning from Sagnac designs~\cite{Steinlechner2017, Vergyris2019,Ecker2021,Lu2023a} to other variants such as periodically poled silica fiber~\cite{Chen2022} or even integrated photonics sources~\cite{Francesconi2023, Miloshevsky2024}. However, 
the integration of controlled unitaries between these two DoFs is essential to maximize the capabilities of this technology. Specifically, in the pursuit of entanglement distillation, the target qubit is often sacrificed (i.e., for postselection) to distill high-quality entanglement from the control qubit~\cite{Hu2021, Ecker2021}. Due to the susceptibility to error accumulation in optical fiber channels for the polarization DoF,  
a single-photon \textsc{cnot} gate with polarization as the control qubit and frequency as the target (which we abbreviate as ``polarization-frequency \textsc{cnot}'' hereafter) will be highly desired. In this context, signal-idler photon pairs hyperentangled in both polarization and frequency can be distributed to, for instance, Alice and Bob through noisy channels. Subsequently, Alice and Bob can each apply a polarization-frequency \textsc{cnot} to the single photon they receive, enabling them to distill a higher quality of polarization entanglement through frequency postselection~\cite{Xu2024}. 

\begin{figure*}[t!]
\centering\includegraphics[width=5in]{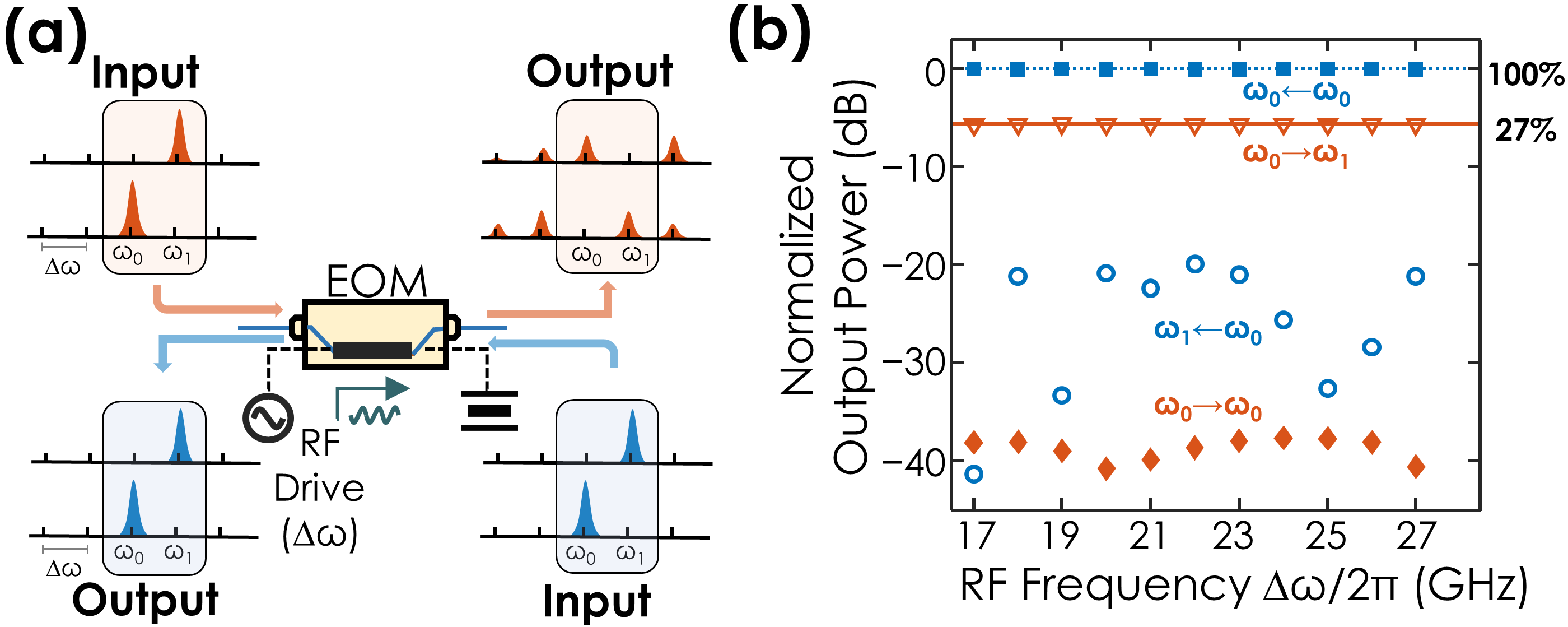}
\caption{(a) Conceptual diagram of unidirectional frequency shifting in a traveling-wave EOM. (b) Frequency conversion efficiencies measured when a single-frequency optical input ($\omega_0$) copropagates (red) and counterpropagates (blue) with the RF signal.}
\label{fig:Concept}
\end{figure*}

In this work, we leverage electro-optic modulation in a fiber Sagnac loop to experimentally demonstrate the first polarization-frequency \textsc{cnot} gate in any photonic platform. 
Leveraging the massive difference in modulation efficiency for copropagating and counterpropagating radio-frequency (RF) and optical fields, our concept furnishes an intrinsically phase-stable and noiseless configuration for polarization-controlled frequency hops. We validate its efficacy through computational basis measurements and synthesis of all four Bell states between polarization and frequency DoFs with fidelities in excess of 98\%. When paired with polarization-frequency hyperentangled sources, our gates should find application in quantum communication protocols 
using this distinctive pair of fiber-compatible DoFs.

\textit{Concept.---}Consider a scenario where a single photon is encoded in two frequency bins ($\omega_0$ and $\omega_1$) and two polarization modes (horizontal $H$ and vertical $V$). In this configuration, the two-qubit computational-basis states can be expressed as: $\ket{H\omega_0}\coloneq\ket{00}$, $\ket{H\omega_1}\coloneq\ket{01}$, $\ket{V\omega_0}\coloneq\ket{10}$, and $\ket{V\omega_0}\coloneq\ket{11}$. A polarization-frequency \textsc{cnot} gate is therefore equivalent to polarization-dependent frequency conversion, where the photon frequency will hop from bin 0 to 1, and vice versa, only when the photon is vertically polarized. Thus, the \textsc{cnot} operation in the two-qubit computational basis can be mathematically described as: $\ket{H\omega_{0(1)}}\xrightarrow[]{\textsc{cnot}} \ket{H\omega_{0(1)}}$, and $\ket{V\omega_{0(1)}}\xrightarrow[]{\textsc{cnot}} \ket{V\omega_{1(0)}}$.

When considering the manipulation of optical frequencies, the primary approach typically involves nonlinear optics, where strong pump(s) can facilitate the conversion of photon frequency between two designated modes through either three-wave or four-wave mixing processes~\cite{Lu2023b}. Additionally, the phase matching condition in most nonlinear media is polarization-sensitive, providing a natural pathway to explore polarization-controlled frequency conversion~\cite{Xu2024}. However, achieving near-unity efficiency remains a significant challenge, and the presence of any unconverted photons in the original frequency bin—--along with added background noise from the optical pump(s)---will inevitably degrade the fidelity of the operation.

On the other hand, the electro-optic phase modulator (EOM) serves as a valuable complement to nonlinear optical methods, particularly for intraband frequency conversion ($<$1~THz) of single photons. Notable demonstrations include the use of a single modulator for spectral shearing~\cite{Wright2017, Zhu2022} and the development of quantum frequency processors---concatenations of EOMs and pulse shapers---for synthesizing quantum frequency gates~\cite{Lukens2017, Lu2019c, Lu2023b}. However, exploring polarization-controlled frequency operations with commercial EOMs presents challenges---most commercial EOMs either accept a single polarization (with integrated polarizer) or accommodate both polarizations, but introduce undesired walkoff and imperfect modulation contrast between the two axes. One feasible approach involves employing a \emph{polarization-diversity} strategy, wherein photons are split into two spatial modes based on their polarization using a polarizing beamsplitter (PBS). Each fiber path then leads to an independent EOM undergoing phase modulation before being recombined at the output through a second PBS~\cite{Sandoval2019}. While this polarization-diverse scheme theoretically should realize the desired \textsc{cnot} gate (e.g., by sending the polarization that will remain unchanged to an undriven EOM), active stabilization between the two optical paths adds to the complexity of this method.

\begin{figure*}[t!]
\centering\includegraphics[width=5in]{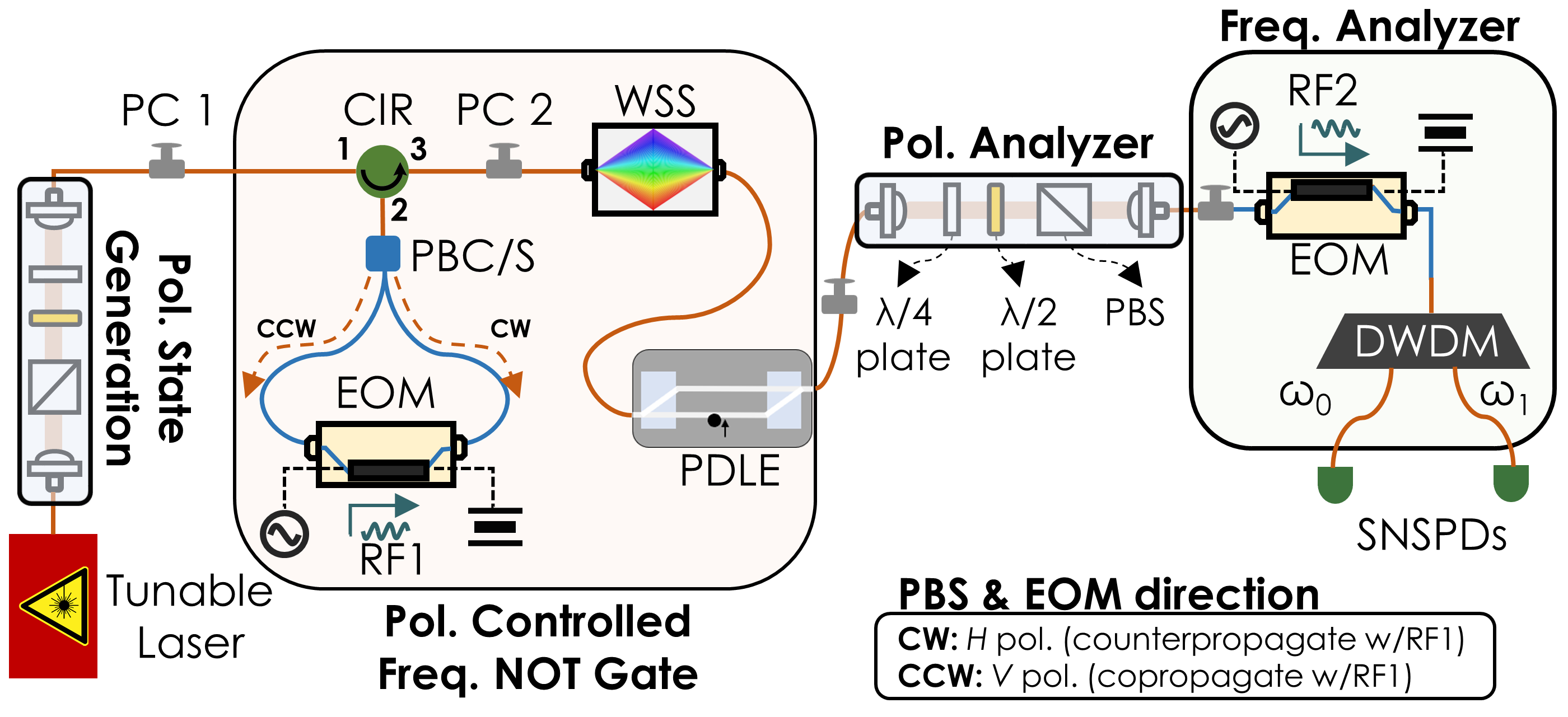}
\caption{Experimental setup for implementation and characterization of polarization-frequency \textsc{cnot} gate. [CIR: fiber-optic circulator. DWDM: 25~GHz dense wavelength-division multiplexer. EOM: electro-optic phase modulator.
PC: polarization controller.
PDLE: polarization-dependent loss emulator. 
PBC/S: fiber-based polarization beam combiner/splitter.
PBS: polarizing beamsplitter cube.
WSS: wavelength-selective switch.
SNSPD: superconducting nanowire single-photon detector.]}
\label{fig:Setup}
\end{figure*}

Here, we introduce an innovative approach that combines the unidirectionality of EOMs with the inherent path stability of the Sagnac interferometer to realize a high-fidelity polarization-frequency \textsc{cnot} gate. Consider a traveling-wave EOM, as illustrated in Fig. \ref{fig:Concept}(a), with RF drives injected from left to right. Optical signals, co-propagating with the RF drive [red arrows from the left in Fig.~\ref{fig:Concept}(a)], accumulate temporal phase modulation, resulting in frequency conversion into newly generated sidebands. Specifically, when we consider a monochromatic optical input (centered at $\omega_0$) and single-sinewave RF modulation (with modulation depth $\Theta$ and frequency $\Delta\omega$), the output spectrum comprises multiple discrete frequency bins spaced by $\Delta\omega$, with their spectral amplitude following a Bessel function of the first kind, $J_k(\Theta)$, where the integer $k$ represents the frequency bin located at $\omega_0+k\Delta\omega$.

When optical signals propagate from the opposite direction [blue arrows in Fig.~\ref{fig:Concept}(a)], however, the temporal phase modulation fails to build up due to the phase mismatch between the two counter-propagating fields, resulting in an extremely low modulation efficiency. Such an effect is particularly pronounced for gigahertz RF drives in off-the-shelf EOMs~\cite{Guo2019, Cardea2023}. These contrasting scenarios provide an intriguing pathway to effectively suppress frequency conversion for one of the polarizations, provided that we can map orthogonal polarizations to co- and counter-propagating directions in the EOM, which will be covered in the following section.

To experimentally verify this proposal, we direct a continuous-wave laser operating at $\omega_0/2\pi=192.0$~THz ($\sim$1561.4 nm) through a fiber-coupled EOM (EOM1; EOSpace). The EOM is driven by pure sinewave modulation, with its amplitude ($\Theta$) and frequency ($\Delta\omega$) set by an RF oscillator (RF1; Agilent), an amplifier, and a variable attenuator. For optimal frequency conversion between $\omega_0$ and $\omega_1$ ($=\omega_0+\Delta\omega$), we select a modulation index $\Theta=2.405$. This choice ensures that the photons at the original bin are depleted, as $J_0(2.405)=0$. However, it also imposes an upper bound on the conversion efficiency to $|J_1(2.405)|^2\approx 27\%$ (excluding the insertion loss of the EOM), with photons scattered to sidebands outside the computational space of our interest (which can be removed with an additional spectral filter). While inherent to the use of a single EOM for frequency shifting~\cite{Lu2023b}, this has minimal impact on the fidelity of the operation given perfect depletion of the original bin---a major distinction from nonlinear frequency conversion methods where nonunity efficiency implies residual energy in the input frequency bin.

Figure~\ref{fig:Concept}(b) summarizes our initial efficiency tests. The red markers denote the optical power in the original bin $\omega_0$ and the converted bin $\omega_1$ (normalized to the total power at the output) when the laser copropagates with the RF signal. We vary the RF frequency from 17~GHz to 27~GHz with 1~GHz increments, within the range predetermined by the RF amplifier. In all cases, the conversion efficiencies closely match the theoretical prediction (red solid line) at 27\%, with an approximate 40~dB depletion of the original bin. Meanwhile, the vast majority of the optical power remains in the original bin for the counterpropagating direction, as indicated by the blue markers. The conversion efficiencies in this direction are all below 1\%, suggesting an effective \emph{counterpropagating} index of $\Theta\approx 0.2$. Together these results imply copropagating (counterpropagating) contrasts of $\gtrsim$40~dB ($\gtrsim$20~dB) between the targeted and suppressed frequency bins. Interestingly, this unidirectionality for EOMs under high-frequency modulation has been leveraged for the realization of optical isolators~\cite{Yu2023}. Moreover, the suppression of modulation for counterpropagating RF and optical fields is much stronger than possible with polarization rotation alone. For example, launching light copropagating with the RF signal but aligned to the fast axis of the EOM fiber pigtail---rather than the slow axis as designed---reduces the modulation index significantly from the slow-axis value of $\Theta=2.405$, but not to zero. Empirically we observe a contrast for copropagating fast-axis light of only $\sim$13~dB, 
appreciably lower than the \emph{counterpropagating slow-axis} value of $\gtrsim$20~dB. Accordingly, the bidirectional approach proves better suited for realizing a high-fidelity polarization-controlled frequency hop, which we employ in the \textsc{cnot} in the subsequent section.

\textit{Polarization-Frequency CNOT Design.---}Figure~\ref{fig:Setup} illustrates the design of our \textsc{cnot} gate. We first connect the two output ports of a fiber-based polarization beam combiner/splitter (PBC/S) to the aforementioned EOM, thereby establishing a fiber Sagnac loop. In this configuration, orthogonal polarizations are mapped to two directions: clockwise (CW) and counterclockwise (CCW), with the former counterpropagating and the latter copropagating with the RF signals. An additional 90-degree rotation is introduced in one of the PBC/S outputs, ensuring that photons in both directions are polarized along the slow axis of the polarization-maintaining fiber in the loop. As both directions also share a common path in the Sagnac interferometer, any imbalance or environmental instabilities are automatically compensated.

Upon recombination at the PBC/S, photons are redirected to the third port of the optical circulator, followed by a wavelength-selective switch (WSS; Finisar) and a programmable polarization-dependent loss emulator (PDLE; OZ Optics). The WSS applies a bandpass filter and blocks any frequency-converted photons that fall outside the qubit subspace (i.e., $\omega_0$ and $\omega_1$). Meanwhile, the PDLE balances the total throughput of the two orthogonal polarizations. Since the CCW path (representing vertical polarization in our convention) experiences more scattering outside of the computational space, we employ a polarization controller (PC2) to align the horizontally polarized inputs (CW path) with the PDL axis and attenuate to 27\%  [5.7~dB; cf. Fig.~\ref{fig:Concept}(b)], thus equalizing the total efficiency of both input polarizations through the \textsc{cnot} circuit. Such additional polarization-selective attenuation is not intrinsic to the method as whole, but required here to match the sideband scattering from a single EOM, as discussed earlier.  
Theoretically, the state transformation matrix $W$ for our \textsc{cnot} design can be expressed as:
\begin{equation}
\label{PolFreq-CNOT}
W = \eta\begin{bmatrix}
1 & 0 & 0 & 0 \\
0 & 1 & 0 & 0 \\
0 & 0 & 0 & -1 \\
0 & 0 & 1 & 0
\end{bmatrix},
\end{equation}
where $\eta$ encapsulates both the insertion loss of the system and the efficiency of the frequency hop. Comparing to the conventional expression, our matrix contains an additional minus sign in the third row, a consequence of our phase convention and properties of the Bessel function, i.e., $\ket{V\omega_{1}}\longrightarrow J_{-1}(\Theta)\ket{V\omega_{0}}$ 
and $J_{-1}(\Theta)=-J_{1}(\Theta)\leq0$ for $\Theta\in[0,3.832]$. 

\begin{figure}[t!]
\centering\includegraphics[clip,trim=1.5cm 0cm 1.5cm 0cm, width=\columnwidth]{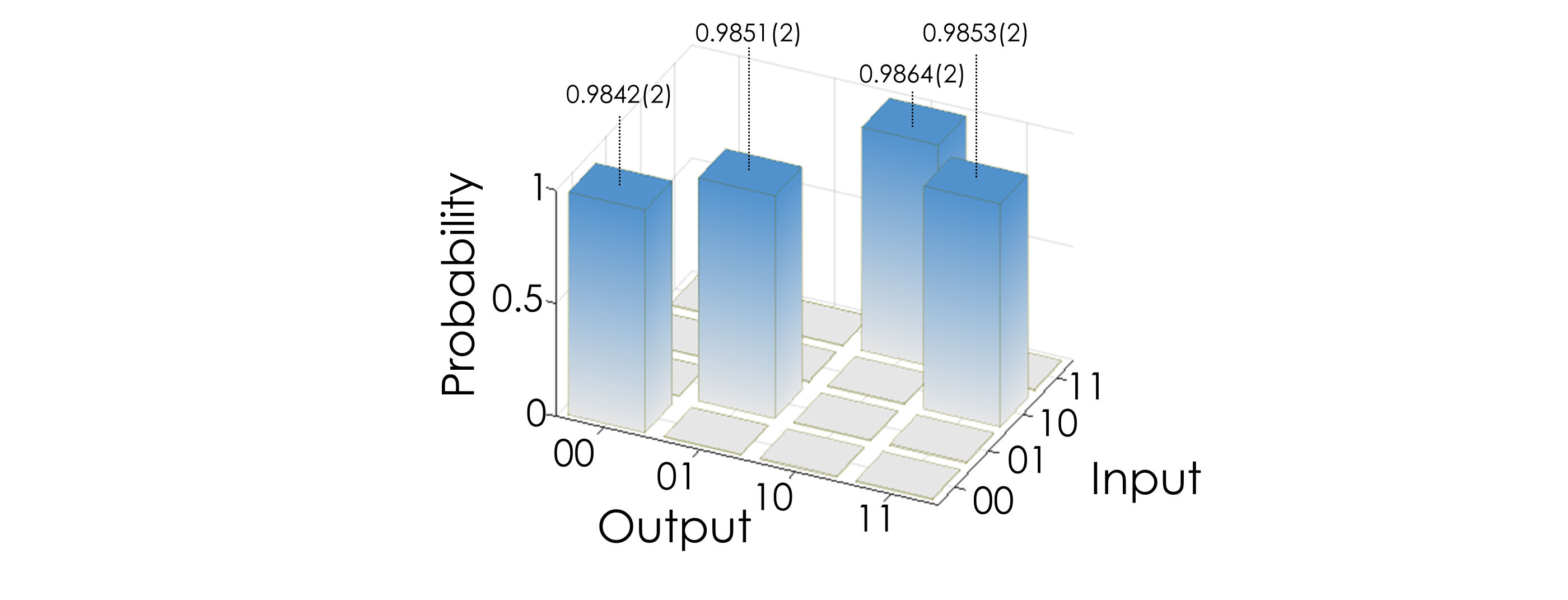}
\caption{Output state probabilities for each computational-basis input state. A characteristic \textsc{cnot} bit flip is observed when the control qubit is in $\ket{1}$ (i.e., polarization is in $\ket{V}$). [State definition: $\ket{H\omega_0}\coloneq\ket{00}$, $\ket{H\omega_1}\coloneq\ket{01}$, $\ket{V\omega_0}\coloneq\ket{10}$, and $\ket{V\omega_0}\coloneq\ket{11}$]}
\label{fig:Comp}
\end{figure}

\textit{Experimental Results.---}To assess the gate's operation, we initiate measurements for both input and output states in the computational basis. We first prepare an input state $\ket{V\omega_0}$, attained through tuning the laser to $\omega_0/2\pi = 192.0$~THz. Vertical polarization is established via the transmitted beam from the free-space PBS with all waveplates aligned to $0^{\circ}$ (cf. Pol. State Generation stage of Fig.~\ref{fig:Setup}). Adjustment of the waveplate angles allows for the preparation of other input polarizations as needed. The EOM within the \textsc{cnot} setup is independently configured to execute a frequency hop operation at frequency spacing of $\Delta\omega/2\pi=25$ GHz. Immediately before the gate input, we employ another polarization controller (PC1) to align the $V$ polarization with the CCW direction, ensuring minimal optical power is retained in the original frequency bin. Since our \textsc{cnot} gate is intrinsically a one-photon operation, we opt to use weak coherent states as input as an accurate proxy for its performance with true single photons~\cite{Lu2018a, Lu2020b}. Consequently, we  attenuate the input laser to single-photon levels, yielding count rates of $\sim$10$^{6}$~s$^{-1}$ in the prepared state.

At the output of the \textsc{cnot} gate, a polarization analyzer is employed, comprising motorized half- and quarter-waveplates, along with a free-space PBS, to execute polarization projections~\cite{Alshowkan2022, Lu2023a}. To characterize the frequency DoF, we incorporate an additional EOM (EOM2), driven by a separate RF oscillator (RF2), which is synchronized with RF1 through a 10~MHz reference. Coupled with a 25~GHz-spaced dense-wavelength division multiplexer (DWDM), this configuration facilitates the realization of all required frequency projections~\cite{Lu2023a, Lu2023b}. The frequency-demultiplexed outputs (i.e., $\omega_0$ and $\omega_1$) are sent to two superconducting nanowire single-photon photodetectors (SNSPDs; Quantum Opus) for detection. 

\begin{figure*}[t!]
\centering\includegraphics[width=5in]{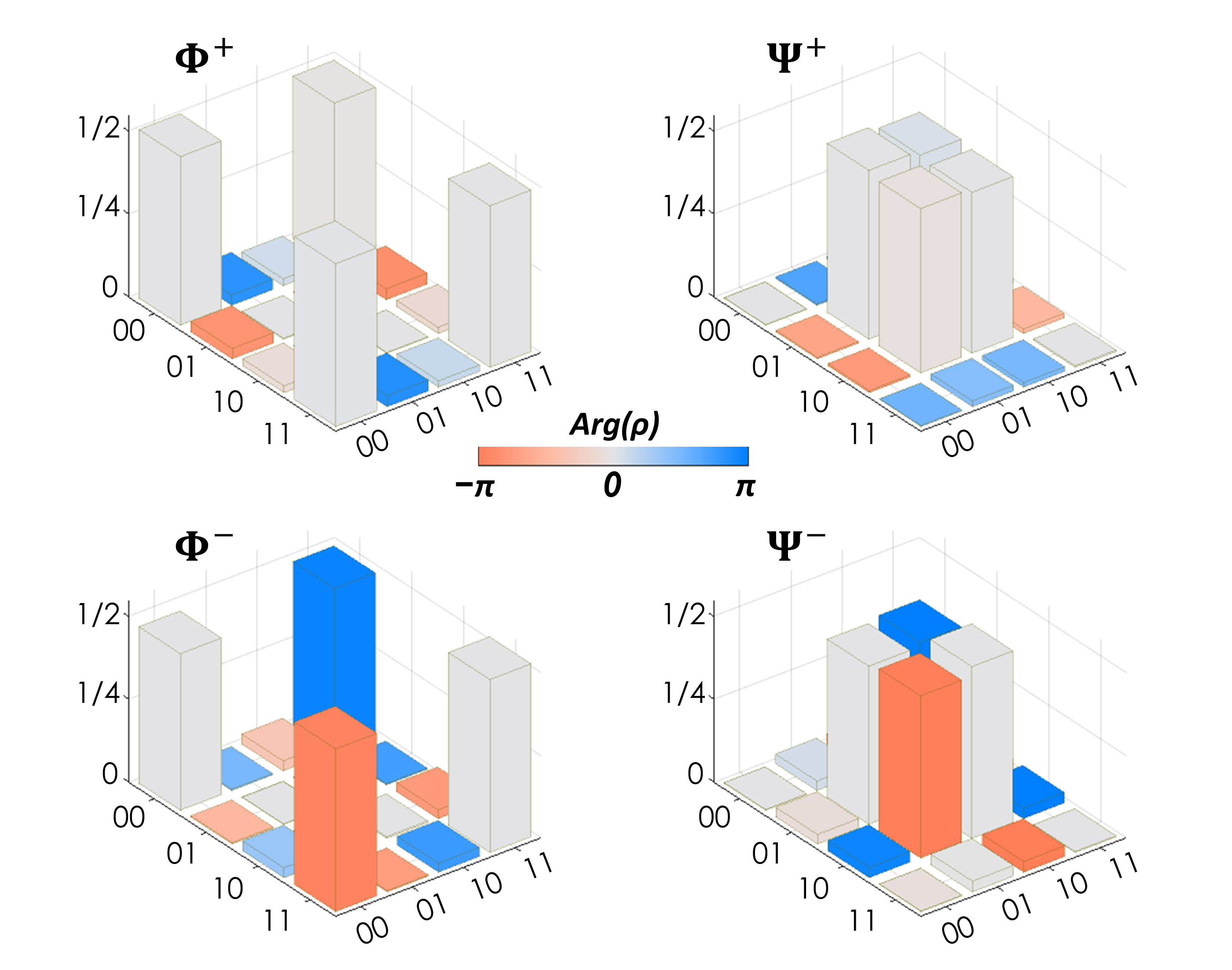}
\caption{Synthesis of all four Bell states with the control (polarization) in a superposition and the target (frequency) in a logical basis state. The measured Bayesian fidelities are $98.8(3)\%$, $98.2(9)\%$, $98.6(3)\%$, and $98.7(3)\%$ for $\ket{\Phi^{\pm}}$ and $\ket{\Psi^{\pm}}$, respectively. [State definition: $\ket{H\omega_0}\coloneq\ket{00}$, $\ket{H\omega_1}\coloneq\ket{01}$, $\ket{V\omega_0}\coloneq\ket{10}$, and $\ket{V\omega_0}\coloneq\ket{11}$] }
\label{fig:Bell}
\end{figure*}

Figure~\ref{fig:Comp} shows the output state probabilities in this truth-table measurement, where the four outcomes for each input state are normalized to sum to unity. When the input features horizontal polarization (control qubit in $\ket{0}$), the output frequency follows the input frequency. Conversely, for vertical polarization (control qubit in $\ket{1}$), the output frequency is flipped as designed, with minimal counts observed in the original frequency bin. Notably, the output polarization faithfully tracks the input in all scenarios. The average probability of obtaining the correct output is $98.53(4)\%$, derived from the mean of the four peaks shown in Fig.~\ref{fig:Comp}. This initial evidence strongly supports the proper operation of our gate, but is not sensitive to the coherence of the computational basis terms, requiring further investigation with superposition inputs.

A quintessential function of the \textsc{cnot} is its ability to synthesize all four maximally entangled Bell states~\cite{OBrien2003}---$\ket{\Phi^{\pm}}=\frac{1}{\sqrt{2}}(\ket{00}\pm\ket{11})$ and $\ket{\Psi^{\pm}}=\frac{1}{\sqrt{2}}(\ket{01}\pm\ket{10})$---when the control qubit (in our example, polarization) is in an appropriate superposition state while the target (frequency) is in a logical basis. To achieve this, we rotate the input polarization to either $\ket{D}=(\ket{H}+\ket{V})/\sqrt{2}$ or $\ket{A}=(\ket{H}-\ket{V})/\sqrt{2}$, and set the frequency to either $\ket{\omega_0}$ or $\ket{\omega_1}$. Ideally, these cases should produce the four polarization-frequency Bell states
\begin{equation}
\begin{aligned}
\label{CNOT}
\frac{1}{\sqrt{2}}(\ket{H}\pm\ket{V})\otimes\ket{\omega_0}\xrightarrow[]{\textsc{cnot}} \frac{1}{\sqrt{2}}(\ket{H\omega_0}\pm\ket{V\omega_1})\coloneq\ket{\Phi^{\pm}}\\
\frac{1}{\sqrt{2}}(\ket{H}\pm\ket{V})\otimes\ket{\omega_1}\xrightarrow[]{\textsc{cnot}} \frac{1}{\sqrt{2}}(\ket{H\omega_1}\mp\ket{V\omega_0})\coloneq\ket{\Psi^{\mp}}
\end{aligned}
\end{equation}
We note that the additional minus sign in Eq.~\ref{PolFreq-CNOT} simply swaps the order of the $\ket{\Psi^\pm}$ states produced compared to a standard \textsc{cnot}. 

We then carry out quantum state tomography, comprising an overcomplete set of 36 projections (6 in each DoF) of the output photons. The collected singles data, integrating over 10 s per projection, are processed through Bayesian inference~\cite{Lukens2020b, Lu2022b}. From a total of 1024 retrieved Bayesian samples, mean density matrices are computed and plotted in Fig.~\ref{fig:Bell}. The fidelity of each case with respect to the ideal Bell states is computed, yielding $\sF=98.8(3)\%$, $98.2(9)\%$, $98.6(3)\%$, and $98.7(3)\%$ for $\ket{\Phi^{\pm}}$ and $\ket{\Psi^{\pm}}$, respectively. The high-fidelity synthesis of all four Bell states confirms the utility of our \textsc{cnot} for coherently ``entangling'' these two DoFs in the formal mathematical sense characteristic of multiqubit single-photon operations.

\textit{Conclusion.---}The reported high-fidelity \textsc{cnot} should facilitate immediate applications such as single-copy entanglement distillation~\cite{Simon2002, Sheng2010, Hu2021, Ecker2021}. In addition, the inherent frequency-translation invariance in the EOM enables straightforward parallelization. Thus, our design has the potential to support entanglement distillation for multiple polarization-entangled channels in parallel when paired with a broadband polarization-frequency hyperentangled source~\cite{Chen2022, Lu2023a}.

A key objective moving forward is to enhance the overall yield of the design by either (i) reducing the insertion loss (presently, $\sim$10 dB) or (ii) improving the frequency conversion efficiency, i.e., the success probability of the \textsc{cnot} gate (capped at 27\%). Concerning (i), the current design's insertion loss is primarily attributable to the EOM ($\sim$3 dB) and the WSS ($\sim$5 dB). However, the function of the WSS can be replaced with a fixed filter where <1~dB should be possible with a 50~GHz DWDM. Regarding (ii), as previously discussed, the conversion efficiency is restricted by the use of a single EOM. This limitation can be addressed with more advanced frequency mixer designs, like the quantum frequency processor~\cite{Lu2020b} and the photonic molecule~\cite{Hu2021b}, both theoretically capable of achieving near-unity frequency conversion but currently hindered by relatively high insertion losses. While further engineering advancements in low-loss integrated modulator designs will be beneficial, our work presents an intriguing platform for intermediate applications, potentially paving the way for even more efficient designs in the future.

\section*{Acknowledgments}
H.H.L. would like to thank J. Hu for insightful discussions that sparked the idea. The authors also extend their appreciation to K. Myilswamy, A. M. Weiner, D. Xu, and L. Qian for valuable discussions. This work was performed in part at Oak Ridge National Laboratory, operated by UT-Battelle for the U.S. Department of energy under contract no. DE-AC05-00OR22725. Funding was provided by the U.S. Department of Energy, Office of Science, Advanced Scientific Computing Research (Field work Proposals ERKJ378, ERKJ353, DE-SC0024257).

%

\end{document}